\documentclass[letterpaper, 10 pt, conference]{ieeeconf}  
\usepackage{graphicx}
\IEEEoverridecommandlockouts                              
\usepackage{amsmath}
\usepackage{amssymb}
\usepackage{graphics}
\usepackage[hidelinks]{hyperref}
\usepackage{xcolor}
\usepackage{dblfloatfix}
\usepackage {tikz}
\usepackage{mathtools}
\usepackage{svg}
\usepackage[bottom]{footmisc}
\usepackage{amsmath}
\usepackage{soul}
\usepackage[switch]{lineno}
\usepackage{multirow}
\usepackage{caption}
\usepackage{tabularray}
\usepackage[linesnumbered,ruled,vlined, noend]{algorithm2e}
\usetikzlibrary {positioning}

\newcommand{\yasin}[1]{\textcolor{blue}{#1}}

\title{\LARGE \bf
Guiding Learning in Multi-Agent Systems using Relational Networks: \\A Value-Based Factorization View
}

\title{\LARGE \bf
Relational Network-guided Learning in Multi-Agent Systems: \\A Value-Based Factorization View
}

\title{\LARGE \bf
Guiding Learning in Multi-Agent Systems using Relational Networks: \\A Value-Based Factorization View
}

\title{\LARGE \bf
How Relational Networks Impact Learning in Multi-Agent Systems: \\A Value-Based Factorization View
}

\title{\LARGE \bf
Emergence of Alternative Behaviors in Multi-Agent Systems using Relational Networks: \\A Value-Based Factorization View
}

\title{\LARGE \bf
Impact of Relational Networks in Multi-Agent Learning: \\ A Value-Based Factorization View
}

\author{Yasin Findik$^{1}$, Paul Robinette$^{2}$, Kshitij Jerath$^{3}$, and  S. Reza Ahmadzadeh$^{1}$
\thanks{$^{1}$ PeARL Lab, Richard Miner School of Computer and Information Sciences, University of Massachusetts Lowell, MA, USA {\tt\small yasin\_findik@student.uml.edu, reza@cs.uml.edu}}%
\thanks{$^{2}$ Department of Electrical and Computer Engineering, University of Massachusetts Lowell, MA, USA {\tt\small paul\_robinette@uml.edu}}%
\thanks{$^{3}$ Department of Mechanical Engineering, University of Massachusetts Lowell, MA, USA {\tt\small kshitij\_jerath@uml.edu}}
}

\begin{document}
\maketitle
\thispagestyle{empty}
\pagestyle{empty}

\begin{abstract}

Effective coordination and cooperation among agents are crucial for accomplishing individual or shared objectives in multi-agent systems. In many real-world multi-agent systems, agents possess varying abilities and constraints, making it necessary to prioritize agents based on their specific properties to ensure successful coordination and cooperation within the team. However, most existing cooperative multi-agent algorithms do not take into account these individual differences, and lack an effective mechanism to guide coordination strategies. We propose a novel multi-agent learning approach that incorporates relationship awareness into value-based factorization methods. Given a relational network, our approach utilizes inter-agents relationships to discover new team behaviors by prioritizing certain agents over other, accounting for differences between them in cooperative tasks. We evaluated the effectiveness of our proposed approach by conducting fifteen experiments in two different environments. The results demonstrate that our proposed algorithm can influence and shape team behavior, guide cooperation strategies, and expedite agent learning. Therefore, our approach shows promise for use in multi-agent systems, especially when agents have diverse properties.

\end{abstract}

\section{Introduction}

Multi-agent scenarios are ubiquitous in various domains such as search and rescue operations~\cite{kleiner2006rfid}, autonomous driving~\cite{pendleton2017perception, Kim2022CACC}, and multiplayer games~\cite{peng2017multiagent}. In such scenarios, the success of achieving shared or individual goals critically depends on the coordination and cooperation between agents~\cite{busoniu2008comprehensive}. Multi-Agent Reinforcement Learning (MARL) approaches have emerged as a popular solution to address the general challenges of cooperation in multi-agent environments. Yet, cooperative MARL faces several challenges that are intrinsic to multi-agent systems, such as the curse of dimensionality~\cite{shoham2007if, Haeri2020Swarm}, non-stationarity~\cite{busoniu2008comprehensive}, and global exploration~\cite{matignon2012independent}. Furthermore, the presence of agents having constraints (e.g., limited battery life, restricted mobility) or distinct roles exacerbates these challenges. 

Among the various MARL methods for cooperative tasks, the Centralized Training with Decentralized Execution (CTDE)~\cite{oliehoek2008optimal} paradigm has become very prominent for addressing numerous cooperative challenges, including but not limited to the ones previously mentioned. While CTDE-based approaches have demonstrated state-of-the-art performance in tackling coordination tasks~\cite{gronauer2022multi}, they lack the ability to steer the coordination strategy that the algorithms converge to. This limitation arises from the fact that these approaches (a) assume agents are identical within the team and (b) do not incorporate any mechanism for specifying the inter-agent relationships. To address this issue, some algorithms have adopted the coordination graph concept~\cite{guestrin2001multiagent} to enable the sharing of action-values among the agents using deep learning~\cite{bohmer2020deep}. However, they require to learn weighing and distributing action-values from a given undirected graph, which may not fully consider the influence of certain behaviors over others. Overall, the problem of agents possessing diverse attributes, a team structure, or a notion of priority among agents remains unresolved.


In this paper, we propose a novel framework that leverages a relational network to capture the relative importance and priorities that agents assign to one another, enabling agents to uncover new cooperation strategies. Our framework uses CTDE paradigm to solve cooperative MARL challenges. We evaluated our approach using fifteen experiments in two distinct environments, each featuring a cooperative task among multiple agents. We have compared our method against the state of the art algorithm, Value Decomposition Networks (VDN)~\cite{sunehag2017value}. Our results and comparisons support three main findings. First, unlike other methods, our approach is highly effective in steering the agents towards a specific team behavior, as governed by the relational network. Second, the experiments reveal that our framework enables agents to uncover new behaviors that promote successful task completion, while also preserving the essential structure of the team. Lastly, our results indicate that the proposed approach accelerates the learning of team behavior, particularly in scenarios where agents face constraints, and relational networks play a critical role.

\section{Related Work}

Multi-Agent Reinforcement Learning (MARL) has gained significant attention in recent years as an active research area, particularly in cooperative settings. Various approaches have been explored to train agents to cooperate effectively towards a shared goal. One such approach is fully centralized learning, where a single controller is shared among all agents, enabling them to learn a joint policy or value function in unison~\cite{claus1998dynamics}. Despite the potential advantages of fully centralized learning, this approach can be computationally expensive and can lead to intractability issues due to the exponential growth of the observation and action space with the increasing number of agents. Alternatively, fully decentralized learning is another approach to MARL in cooperative settings, where each agent independently learns its own policy. In such approaches, the cooperative behavior arises from the learned policies as they are applied in the environment. For instance, Independent Q-Learning (IQL)~\cite{tan1993multi} trains each agent using a separate action-value table via Q-learning~\cite{watkins1992q}. Since tabular Q-learning is insufficient for addressing high-dimensional state and action space, IQL framework was later extended with function approximation~\cite{tampuu2017multiagent}. Yet, independent learning in multi-agent settings is susceptible to non-stationarity, stemming from other agents' actions from the perspective of one agent. Due to the invalidity of the Markov property in non-stationary environments, the convergence of decentralized algorithms based on Q-learning cannot be ensured~\cite{hernandez2017survey}. Later, the introduction of fully decentralized algorithms featuring networked agents~\cite{zhang2018fully} capable of communicating and exchanging information has emerged as a promising approach for addressing the non-stationarity problem. In such frameworks, the connections between agents are governed by an undirected graph. In our approach, we also utilize a (directed) graph to represent the rapport among the agents. However, in contrast to frameworks that prioritize the use of a communication channel~\cite{zhang2018fully}, our graph serves to symbolize and establish relations between agents, resulting in a relational network that directly influences team behavior.

To overcome the limitations associated with fully centralized and fully decentralized learning in cooperative MARL scenarios, a novel approach has been proposed known as Centralized Training with Decentralized Execution (CTDE)~\cite{oliehoek2008optimal}. This paradigm enables individual agents to execute their actions while a centralized mechanism integrates their strategies, thereby ensuring effective coordination and alignment towards a shared goal. The CTDE paradigm has been implemented using both policy-based and value-based methods. Specifically, policy-based methods, such as Multi-Agent Deep Deterministic Policy Gradient (MADDPG)~\cite{lowe2017multi} and Multi-Agent Proximal Policy Optimization (MAPPO)~\cite{yu2021surprising}, incorporate a critic that can take into account the global observations of all agents. On the other hand, value-based techniques, including Value Decomposition Networks (VDN)~\cite{sunehag2017value}, QMIX~\cite{rashid2020monotonic}, and QTRAN~\cite{son2019qtran}, enhance Q-Learning by integrating a centralized function that calculates the joint Q-value based on the individual action-values of each agent. These techniques have been shown to be effective in addressing the challenges of multi-agent coordination and achieving higher performance in various scenarios.

Prior research in MARL has predominantly concentrated on achieving an optimal solution for cooperation problems, ranging from fully centralized learning to CTDE paradigm. Yet, the attainment of an optimal or sub-optimal solution by a multi-agent team is intrinsically contingent upon the underlying dynamics of interaction and collaboration among 
agents~\cite{ baker2020emergent, haeri2022reward}. 
While they utilize reward sharing among agents to adapt individual rewards based on their relationships represented as a relational network, our method involves adjusting the agents' contribution to the team reward. This adjustment is specifically determined by the relational network and does not rely on reward sharing among agents. Ultimately, our method ensures that agents still receive the same individual reward provided by the environment.

Our study proposes a framework that allows for the integration of relational networks into the CTDE approaches to gain a deeper understanding of the impact of inter-agent relationships on cooperation strategies and team interactions. Specifically, our investigation focuses on the context of value-based factorization. However, it is worth noting that the use of relational networks, as proposed in our study, can be extended to other CTDE algorithms.

\section{Background}

\subsection{Markov Decision Process}
\label{MDP}
We characterized Decentralized Markov Decision Process as a tuple $\langle  \mathcal{S}, \mathcal{A}, \mathcal{R}, \mathcal{T}, \gamma\rangle$ where $s \in \mathcal{S}$ indicates the true state of the environment, the joint set of individual actions and rewards are represented by $\mathcal{A} \coloneqq \{a_1, a_2, \dots, a_n \}$, $\mathcal{R} \coloneqq \{r_1, r_2, \dots, r_n \}$, respectively,  $\mathcal{T} (s, A, s') \colon \mathcal{S} \times \mathcal{A}\times \mathcal{S} \mapsto [1,0]$ is the dynamics function defining the transition probability, and $\gamma\in[0,1)$ is the discount factor.

\subsection{Deep Q-Learning}
The Deep Q-Network (DQN)~\cite{mnih2015human} is a variant of an action-value method, which defines $Q(s, a)=\mathbb{E}[G | S=s, A=a]$ where $G$ denotes the return, that employs a deep neural network to approximate the Q-function, represented as $\hat{Q}(s, a, \theta)$ where $\theta$ is the weight vector. Unlike tabular Q-learning, DQN is capable of effectively handling high-dimensional state and action spaces by leveraging deep learning. However, the training of DQN poses a significant challenge due to instability and divergence caused by the Q-network parameters being updated in each step, which violates the assumption of independently and identically distributed (i.i.d) data points. To address these challenges, Mnih et al.~\cite{mnih2015human} have also proposed several techniques, including experience replay and fixed Q-target networks, which have since become standard in many deep reinforcement learning algorithms.

In a nutshell, two deep Q-neural networks are used for each Q-function, namely: Prediction Neural Network (P-NN), and fixed Target Neural Network (T-NN) which is the P-NN from a previous iteration. Also, a replay memory is utilized to keep a large number of transitions experienced by the agent during its interactions with the environment. Each transition consists of a tuple $\langle s, a, r, s'\rangle$. To train the P-NN, a batch of transitions of size $b$ is sampled from memory. The Temporal Difference (TD) error is calculated between the P-NN value and T-NN value, as follows:
\begin{align}
\label{td_error}
e_{\textrm{TD}} = \sum_{i=1}^{b} [r + \gamma \max_{a'}(Q(s', a', \theta_{t})) - Q(s, a, \theta_{p})], 
\end{align}
where $\theta_p$ are the weights of the P-NN, and $\theta_t$ are those of the T-NN which is regularly updated with $\theta_p$. The parameters of the P-NN are updated using an optimizer in the direction of reducing $e_\textrm{TD}$. In centralized multi-agent systems, a single Q-function is employed for the meta-agent, providing access to the observations of all agents, whereas, in fully decentralized systems, each agent learns its Q-function independently using DQN.

\subsection{Value Function Factorization}

Value function factorization methods, which adhere to the CTDE paradigm, have been proposed as an efficient approach for handling a joint action-value function, whose complexity increases exponentially with the number of agents. These methods effectively address the non-stationarity problem of decentralized learning through centralized training and overcome the scalability issue of centralized learning through decentralized execution. VDN~\cite{sunehag2017value} and QMIX~\cite{rashid2020monotonic} are two exemplary methods for factorizing value functions.

VDN and QMIX both maintain a separate action-value function $Q$ for each agent $i \in \{ 1,...,n\}$. They merge these individual $Q_i$ values to obtain the central action value $Q_{\textrm{tot}}$ using additivity and monotonicity, respectively. Specifically, VDN sums $Q_i$s to obtain $Q_{\textrm{tot}}$, as 
$$Q_{\textrm{tot}} = \sum_{i=1}^{n} Q_i(s, a_i),$$
while QMIX combines them using a state-dependent continuous monotonic function, as follows:
$$ Q_{\textrm{tot}} = f_s(Q_1(s, a_1), ..., Q_n(s, a_n)), $$
where $\frac{\partial f_s}{\partial Q_i} \ge
0,  \forall i \in \{1, ..., n\}$.

These value function factorization methods most commonly utilize DQN to approximate the action value function. Different from decentralized learning where all agents' individual P-NNs are trained independently by using~\eqref{td_error}, or centralized learning where a single P-NN exists and is trained with~\eqref{td_error} by providing access to the observations of all agents, in these value function factorization methods, $Q_{\textrm{tot}}$ is used to compute $e_\textrm{TD}$ by updating~\eqref{td_error} as follows:
\begin{align}
\label{td_error_vdn}
e_{\textrm{TD}} = \sum_{i=1}^{b} [r_{\textrm{team}} + \gamma \max_{u'}(Q_{\textrm{tot}}(s', u', \theta_{t})) - Q_{\textrm{tot}}(s, u, \theta_{p})], 
\end{align}
where $r_{\textrm{team}}$ is the sum of agents' rewards with uniform weights, $u$ is the joint action of the agents. The agents' P-NN are trained using an optimizer in the direction reducing obtained $e_\textrm{TD}$. The coordination of agent actions towards maximizing the team reward is facilitated by this process. Consequently, the central aspect of the CTDE paradigm becomes apparent: the agent networks are trained using a centralized $Q_{\textrm{tot}}$, while the actions of each agent are determined by its own neural network, rendering the execution decentralized.

\section{Proposed Method}

\begin{figure}[t]
\centering
  \includegraphics[width=\linewidth]{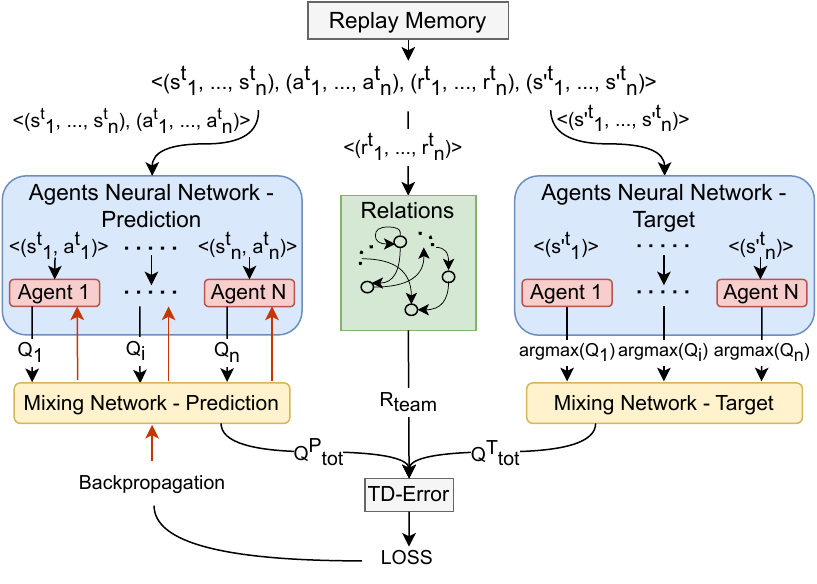}
  \caption{\small{The overall relationship awareness framework. In blue are the agents' P-NNs and T-NNs. The relational network is represented in green. Mixing networks are used combine the individual action-values to obtain the central action-value.}} 
    \label{fig:system_architecture}
\end{figure}

In cooperative MARL, the presence different team structures often yield multiple solutions with varying optimality. Value factorization methods and others, strive to maximize team rewards and converge towards one of several solutions, possibly achieving the global optimum. However, stochasticity in agents' exploration can influence convergence towards a particular team behavior if multiple cooperation strategies exist with the same maximum total reward. Furthermore, in real-world settings, individual agents/robots may be subject to varying constraints or possess distinctive attributes that significantly influence overall team behavior, making it challenging to learn a viable solution without a deeper understanding of the team structure. We propose a novel framework, depicted in Fig.~\ref{fig:system_architecture}, that incorporates relationship awareness into the agents' learning algorithm. And, we opt to explore and study the framework using VDN, as it is a simple yet effective CTDE approach for learning cooperative behaviors, leading us to introduce Relationship-Aware VDN (RA-VDN). RA-VDN is a generalization over VDN that enhances team behavior, provides control over optimal solutions, and expedites agent learning by utilizing the relationships between agents.

The inter-agent relationships are represented by a relational network which is utilized to integrate the individual rewards to obtain the team reward used in~\eqref{td_error_vdn}. The MDP, defined in~\ref{MDP}, is modified to include the relational network which is characterized by a directed graph $\mathcal{G}=(\mathcal{V}, \mathcal{E}, \mathcal{W})$ where each agent $i \in \{ 1,...,n\}$ is a vertex $v_i$, $\mathcal{E}$ is the set of directed edges $e_{ij}$ directed from $v_i$ to $v_j$, and the weight of the edges is represented by the matrix $\mathcal{W}$ with elements $w_{ij}\in[0, 1]$ for each edge. The direction and weight of each edge signify the relationship between the agents. An edge $e_{ij}$ directed from $i$ to $j$ indicates that agent $i$ cares about or is vested in the outcomes for agent $j$. Based on the modified MDP represented as a tuple $\langle  \mathcal{S}, \mathcal{A}, \mathcal{R}, \mathcal{T}, \mathcal{G}, \gamma\rangle$, the team reward defined as follows:
\begin{align*}
\label{reward}
r_{\textrm{team}} = \sum_{i\in\mathcal{V}}^{} \sum_{j\in\mathcal{E}_i}^{} w_{ij}r_j,
\end{align*}
where $\mathcal{E}_i$ denotes the set of vertex indices that have an edge directed from $v_i$, and $r_j$ is the reward of the agent represented by $v_j$. For example, the directed graph as follows:
\begin{center}
\begin {tikzpicture}[-latex,auto ,node distance =3 cm and 2.5cm ,on grid , semithick, state/.style ={ circle, top color=white ,  draw, black, bottom color = black!20, text=black, minimum width =0.5 cm}]
\node[state] (A) []  {$A1$};
\node[state] (B) [right =of A] {$A2$};
\path (A) edge [loop left] node[left] {$w_{11}=1$} (A);
\path (B) edge [loop right] node[right] {$w_{22}=1$} (B);
\path (A) edge node[above =0.01 cm] {$w_{12}=0.5$} (B);
\end{tikzpicture}
$$R_\textrm{team}=r_1 + 1.5r_2$$
\end{center}
illustrates the relationships between two agents. Examining the team reward obtained using this relational network reveals that the second agent's contribution to the team reward is increased by adding half of the reward second agent receives with the assistance of first agent. By giving more emphasis on the second agents, the algorithm can produce various emergent behaviors, such as the first agent helping the second agent in reaching its goal, or sacrificing own individual reward entirely to increase the second agent's reward. Furthermore, it can accelerates the learning of team behavior by agents.

Similar to other CTDE approaches, in VDN, the primary objective of each agent $i$ is to obtain a policy, that optimizes the team performance. The learned policy is then employed to determine the optimal action of the agent $a_i$ for a given state $s$, as follows:
$$a_i = \pi^{\textrm{VDN}}_i(s) = \max_{a}(Q_i(s, a, \theta_{p})), $$
where $\pi_i$ represents the policy of agent $i$, and $Q_i$ is represented using DQN to approximate its action-value function. For all agents, the notation takes the following form:
$$u = \pi^{\textrm{VDN}}(s) = \max_{u}(Q(s, u, \theta_{p})), $$
where $u$ is a vector, containing the actions of all the agents and $\pi$ is a vector of policies, where each element $\pi_i$ represents the policy of the respective agent $i$. We formulate RA-VDN, as follows:
$$u = \pi^{\textrm{RA-VDN}}(s|\mathcal{G}) = \max_{u}(Q(s, u, \theta_{p}|\mathcal{G})), $$
where the policy is conditioned by the relational network $\mathcal{G}$. 

We would like to highlight that in cases where $\mathcal{G}$ represents a self-interest relational network the policies for RA-VDN and VDN become equivalent. That is because in VDN, each agent contributes their reward equally to the team reward.
$$\pi^{\textrm{RA-VDN}}(s|\mathcal{G}) \equiv \pi^{\textrm{VDN}}(s),$$
where $\mathcal{G} \in \{ \textrm{a}, \textrm{d}, \textrm{h} \}$ networks depicted in Fig.~\ref{fig:relations}. Overall, RA-VDN allows for the shaping and influencing of team behavior among agents utilizing different relational networks, whereas VDN leads to agents randomly converging to one of multiple solutions without any control over the cooperation strategies. Also, noted from the experiments, it accelerates convergence of the agents to the team behavior where agents have constraints.

\section{Experiments}

\subsection{Environments}
\label{env}
To assess the efficacy of the proposed approach in shaping and enhancing team behaviors, we applied the RA-VDN algorithm to two environments: a multi-agent grid-world environment, and the \textit{Switch} environment proposed in~\cite{magym}.

\begin{figure}[t]
\centering
  \includegraphics[width=\linewidth]{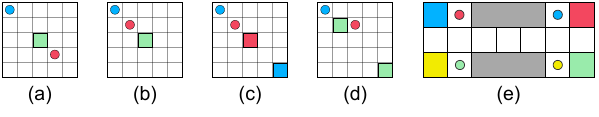}
  \caption{\small{(a-d) multi-agent grid-world environment. (e) \textit{Switch} environment with four agents. In the three-agent setup, the yellow agent and its goal location were removed. For the two-agent setup, the green agent and its goal location were in addition removed.
  }}
  \label{fig:environments}
\end{figure}

\subsubsection{Multi-agent Grid-world Environment}

We created a $5\textrm{x}5$ grid-world environment with resources and two agents, colored red and blue (see Fig.~\ref{fig:environments}(a-d)). The resources are categorized based on their color, green resources are consumable by both agents, while red and blue resources can only be consumed by their respective colored agents. The objective of each episode is for the agents to consume all the resources by visiting their locations, with each agent limited to one resource per episode. The agents have five possible actions: move up, down, left, right, or stay still. They can also \textit{push} each other if they are adjacent and the pushing agent takes a non-idle action towards the pushed agent, which must be idle. After a \textit{push}, the pushing agent remains in place while the other agent moves one space in the pushed direction. Consuming a resource yields $+10$ reward, while the agents are penalized with $-1$ for each time step per unconsumed resource unless they occupy a resource location which serves as a safe spot. The episode ends when all resources are consumed or the maximum time steps are reached. We designed this environment to challenge agents with real-world constraints, such as limited battery life and restricted mobility, and to demonstrate how relational networks can help achieve cooperative tasks by prioritizing based on the agents' constraints.

\subsubsection{Switch Environment}

Figure~\ref{fig:environments}(e) illustrates the \textit{Switch} environment, which comprises a grid-world divided into two regions connected by a narrow bridge. The environment accommodates up to four agents with distinct colors, each aiming to reach their respective goal boxes of matching color on the opposite end. This is challenging due to the narrow bridge, which allows only one agent to cross at a time. Agents can move left, right, up, down, or stay still, and receive a reward of $+5$ upon reaching their goal, but incur a penalty of $-0.1$ per time step away from it. The episode ends after $50$ time steps or when all agents reach their goals, requiring collaborative efforts among agents to maximize rewards and prevent obstructing each other's paths. We select this environment as it offers various cooperation strategies and facilitates evaluating relational networks. Additionally, it was employed in the original VDN paper, enabling a lucid comparison between our approach and VDN.

\subsection{Models and Hyperparameters}

In setting our algorithm up for both environments, we employed a Multi-Layer Perceptron (MLP) with two hidden layers, each consisting of $128$ neurons and utilizing the ReLU activation function. The prediction model for each agent is trained $10$ times per episode using uniformly drawn batches of size $b=32$ selected from the replay memory, which has a capacity of $50$k time-steps. We used \textit{Adam} optimizer with a learning rate of $0.001$ and the loss function is the squared TD-error. The weights of the target network are updated with those of the prediction network every $200$ episodes. To promote exploration, an $\varepsilon$-greedy method is applied, where $\varepsilon$ is linearly annealed over time. Finally, the discount factor, $\gamma$ was set to $0.99$.

The results of the experiments for both environments, presented in Fig.~\ref{fig:results_custom} and~\ref{fig:results_switch}, illustrate the average training reward over 10 runs as represented by the shaded regions, and the average test rewards of the agents, denoted by the solid lines. The test rewards were determined by evaluating the individual rewards of agents based on a greedy strategy, with the training process being interrupted every 50 episodes.

\subsection{Results for Multi-agent Grid-world Environment}
\label{custom_result_text}

We evaluated the performance of RA-VDN through eight independent experiments across four distinct configurations utilizing various relational networks within this environment, and compare the individual rewards obtained by the agents.

\begin{figure}[t]
\centering
  \includegraphics[width=\linewidth]{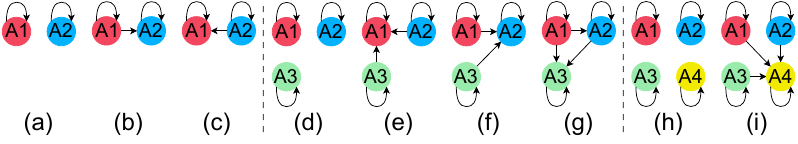}
  \caption{\small{
  Relational networks employed in RA-VDN.}} 
  \label{fig:relations}
\end{figure}

\begin{figure*}[t]
\centering
  \includegraphics[width=\linewidth]{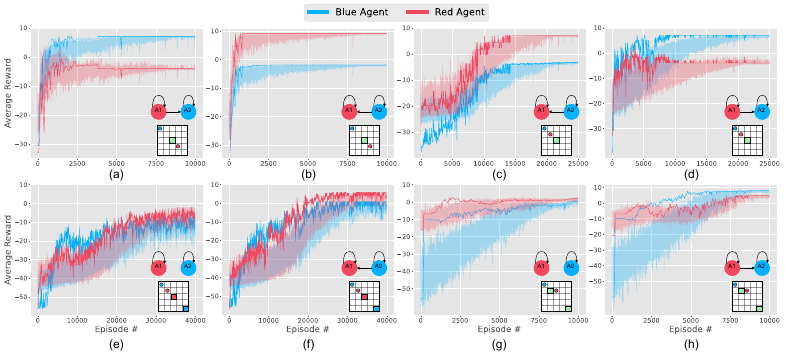}
    \caption{\small{Multi-agent grid-world environment results with different relational networks (details in Section~\ref{custom_result_text}).}}
  \label{fig:results_custom}
\end{figure*}

\vspace{0.3em}
\noindent\textbf{Resource Collection (RC) Scenario:}
This scenario comprises of two agents, represented by red and blue circles and an undedicated resource depicted by a green rectangle, as shown in Fig.~\ref{fig:environments}(a). The resource has intentionally been positioned in closer proximity to the red agent. The determination of which agent is going to consume the given resource is contingent upon a relative importance of the agents within the team. When the importance of agents is equal (like in VDN), as in Fig.~\ref{fig:relations}(a), or when the importance of the red agent surpasses that of the blue agent, as in Fig.~\ref{fig:relations}(c), the red agent consumes the resource since 
resource is located closer to it (see Fig.~\ref{fig:results_custom}(b)). Nevertheless, by introducing a relational network that assigns greater importance to the blue agent (Fig.~\ref{fig:relations}(b)), the red agent reserves the resource for the blue agent's consumption, as depicted in Fig.~\ref{fig:results_custom}(a). By enabling the value decomposition algorithm to access the relationships among agents, it becomes feasible for agents to discover novel behaviors.

\vspace{0.3em}
\noindent\textbf{Resource Collection with Restricted Mobility (RC-RM) Scenario:} 
As illustrated in Fig.~\ref{fig:environments}(b), this experiment involves two agents, with the red agent restricted to vertical movement, and one resource positioned out of reach for the red agent without the help of blue. After training for $25000$ episodes using the relationship in Fig.~\ref{fig:relations}(c), where the red agent's significance surpasses the blue agent's, the blue agent assisted the red agent in reaching the resource, resulting in a higher individual reward for the red agent ($7.0\pm0.0$) than the blue agent ($-3.0\pm0.3$), as shown in Fig.~\ref{fig:results_custom}(c) and Table~\ref{results_custom_table}. 
However, when the relational network in Fig.~\ref{fig:relations}(b) is employed, the red agent deviates from the blue agent's path to the resource, resulting in the blue agent ($7.0\pm0.0$) receiving a higher reward than the red agent ($-4.0\pm0.0$), as shown in Fig.~\ref{fig:results_custom}(d). These findings show the significance of the relative importance of agents in determining their behavior and outcomes in team settings.

\begin{table}[b]
\centering
\caption{\small{Average reward with 95\% confidence intervals for ten runs on multi-agent grid environment after training completed.}}
\resizebox{\linewidth}{!}{%
\begin{tblr}{
  width = \linewidth,
  colspec = {Q[200]Q[100]Q[100]Q[100]Q[100]Q[150]Q[100]Q[100]Q[100]},
  cells = {c},
  cell{1}{2} = {c=2}{0.17\linewidth},
  cell{1}{4} = {c=2}{0.17\linewidth},
  cell{1}{6} = {c=2}{0.17\linewidth},
  cell{1}{8} = {c=2}{0.17\linewidth},
  vline{3,5,7} = {1}{},
  vline{4,6,8} = {1-4}{},
  hline{1,5} = {-}{},
  hline{2} = {2-9}{},
}
                   & {RC\\Scenario} &             & RC-RM Scenario &             & DRC-RM Scenario &             & RC-BC Scenario &             \\
Relational Network & Fig. ~\ref{fig:relations}(b)      & Fig. ~\ref{fig:relations}(c)   & Fig. ~\ref{fig:relations}(c)      & Fig. ~\ref{fig:relations}(b)   & Fig. ~\ref{fig:relations}(a)       & Fig. ~\ref{fig:relations}(c)   & Fig. ~\ref{fig:relations}(a)      & Fig. ~\ref{fig:relations}(b)   \\
Red Agent          & -4.2 ± 0.4    & 9.0 ± 0.0 & 7.0 ± 0.0    & -4.0 ± 0.0 & -7.4 ± 11.1     & 2.4 ± 6.3 & 2.0 ± 1.0    & 5.0 ± 0.0 \\
Blue Agent         & 7.0 ± 0.0    & -2.0 ± 0.0 & -3.0 ± 0.3    & 7.0 ± 0.0 & -11.7 ± 11.6     & 0.9 ± 3.1 & 0.3 ± 2.3    & 7.9 ± 0.2 
\end{tblr}
}
\label{results_custom_table}
\end{table}

\vspace{0.3em}
\noindent\textbf{Dedicated Resource Collection with Restricted Mobility (DRC-RM) Scenario:} 
In this setup, the environment features two agents and two resources, with each resource dedicated to an agent of the same color, as shown in Fig.~\ref{fig:environments}(c). Similar to the prior experiment, the red agent can only move vertically. The optimal solution entails the blue agent helping the red agent before consuming its assigned resource and only then proceeding to consume its own resource. The findings of training with VDN for $40000$ episodes show that the blue agent only learns the optimal solution in $2$ out of $10$ runs ($20$\%), resulting in the blue agent's reward being ($-11.7\pm11.6$) and the red agent's reward being ($-7.4\pm11.1$), as depicted in Table~\ref{results_custom_table}. Yet, when using RA-VDN, the relational network in Fig.~\ref{fig:relations}(c) accelerates the agent learning process, leading to significantly higher rewards for both the red agent ($2.4\pm6.3$) and blue agent ($0.9\pm3.1$), as shown in Fig.~\ref{fig:results_custom}(f) and Table~\ref{results_custom_table}. RA-VDN outperformed VDN by achieving the optimal solution in $80$\% of runs and yielding higher agent rewards. Overall, this experiment demonstrates how proper relationships can expedite the learning of team behavior when any of the agents need special assistance, and RA-VDN benefits from it.

\vspace{0.3em}
\noindent\textbf{Resource Collection with Battery Constraint (RC-BC) Scenario:} 
In this experiment, two agents with battery constraints and two undedicated resources are used, as shown in Fig.~\ref{fig:environments}(d). The red and blue agents have a maximum limit of $10$ and $5$ non-idle actions per episode, respectively, to mimic battery constraints. One resource is located beyond the blue agent's reach due to its limited battery, making the optimal policy for the team to have the red agent to consume the distant resource and leave the nearest one for the blue agent. However, after training VDN for over $10000$ episodes, the red agent failed to learn this strategy, as indicated in Fig.~\ref{fig:results_custom}(g). It learned to relinquish the nearest resource for the blue agent to use as a safe spot after consumption. Unfortunately, this led to the other resource remaining unconsumed since an agent could consume only one resource, causing episode termination solely due to the time-step threshold. Yet, by employing RA-VDN with the relational network illustrated in Fig.~\ref{fig:relations}(b), the red agent learns to reserve the nearest resource for the blue agent's consumption and proceed towards the other resource as it cares about the blue agent. As shown in Fig.~\ref{fig:results_custom}(h) and Table~\ref{results_custom_table}, the blue agent receives an individual reward of ($7.9\pm0.2$), while the red agent gets ($5.0\pm0.0$). These results indicate that RA-VDN is capable of finding and converging to the optimal policy, while VDN fails to resolve the environment where agents have different battery life.

\subsection{Results for Switch Environment}
\label{switch_result_text}
\begin{figure*}[t]
\centering
  \includegraphics[width=\linewidth]{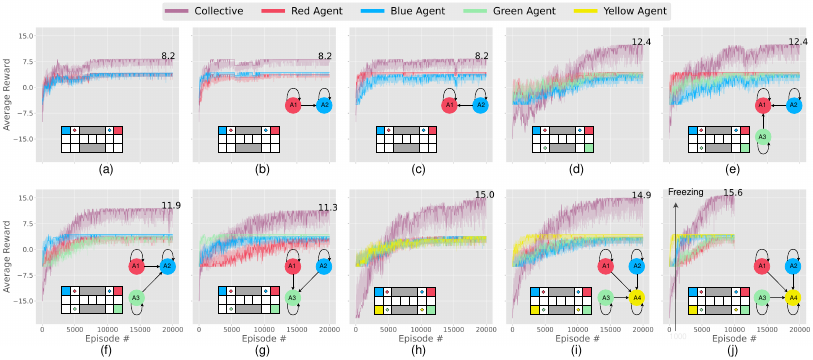}
    \caption{\small{\textit{Switch} environment results with varying number of agents and different relational networks (details in Section~\ref{switch_result_text}).}}
  \label{fig:results_switch}
\end{figure*}

We conducted seven independent experiments in the \textit{Switch} environment using two, three, or four agents. Our algorithm was compared against VDN based on both individual and collected rewards by the team. In Fig.~\ref{fig:results_switch}, collective reward shows the sum of individual rewards attained by the agents (i.e., the team reward maximized in VDN). Note that collective reward differs from team reward used to train RA-VDN, which is computed using a relational network.

\begin{table*}[b]
\centering
\caption{
\small{Average reward with 95\% confidence intervals for ten runs on \textit{Switch} environment after training completed.
}}

\resizebox{\textwidth}{!}{
\begin{tabular}{cccc|cccc|ccc}
\hline
                                       & \multicolumn{3}{c|}{Two-Agent Scenario}   & \multicolumn{4}{c|}{Three-Agent Scenario}     & \multicolumn{3}{c}{Four-Agent Scenario} \\ \cline{2-11} 
                                       & VDN         & \multicolumn{2}{c|}{RA-VDN} & VDN       & \multicolumn{3}{c|}{RA-VDN}       & VDN        & RA-VDN     & Frozen RA-VDN \\ \cline{2-11} 
\multicolumn{1}{l}{Relational Network} & N/A         & Fig.~\ref{fig:relations}(b)          & Fig.~\ref{fig:relations}(c)         & N/A       & Fig.~\ref{fig:relations}(e)       & Fig.~\ref{fig:relations}(f)       & Fig.~\ref{fig:relations}(g)       & N/A                & Fig.~\ref{fig:relations}(i)& Fig.~\ref{fig:relations}(i)           \\
Red Agent                              & 4.10±0.19   & 3.80±0.00    & 4.39±0.02    & 4.38±0.04 & 4.40±0.00 & 3.75±0.03 & 3.13±0.03 & 3.70±0.34  & 3.43±0.29  & 3.47±0.15           \\
Blue Agent                             & 4.15±0.15   & 4.40±0.00    & 3.86±0.06    & 3.66±0.04 & 3.70±0.00 & 4.40±0.00 & 3.78±0.02 & 3.78±0.25  & 3.61±0.17  & 4.20±0.00           \\
Green Agent                            & ---         & ---          & ---          & 4.35±0.02 & 4.30±0.00 & 3.76±0.07 & 4.40±0.00 & 3.77±0.36  & 3.54±0.23  & 3.54±0.14           \\
Yellow Agent                           & ---         & ---          & ---          & ---       & ---       & ---       & ---       & 3.80±0.30  & 4.34±0.09  & 4.40±0.00          \\ \hline
\end{tabular}
}
\label{results_switch_table}
\end{table*}

\vspace{0.3em}
\noindent\textbf{Two-Agent Scenario:}
As depicted in Fig.~\ref{fig:environments}(e), this scenario includes a red and a blue agent with their corresponding goal locations that marked with the same color. Given that only one agent can cross the bridge at a time, there are two optimal policies for this scenario: either the red agent crosses the bridge ahead of the blue agent or the blue agent does. VDN, in each run, converges to one of the two optimal policies because of the algorithm's equal weighting of both agents (i.e., this is the equivalent of the network in Fig.~\ref{fig:relations}(a)), which leads to a randomized selection of optimal policies. As shown in Fig.~\ref{fig:results_switch}(a) and Table~\ref{results_switch_table}, the individual rewards for both agents are nearly identical as a result of averaging their rewards over $10$ runs. In other words, in each run the agent that crosses the bridge maximizes its individual reward and the average value for both agents become similar as the number of runs increases. For RA-VDN, we introduce two relational networks: first the red agent places importance to the blue agent (Fig.~\ref{fig:relations}(b)); second, the blue agent assigns importance to the red agent (Fig.~\ref{fig:relations}(c)). These networks allow us to investigate team behavior and solution preference. The results in Fig.~\ref{fig:results_switch}(b) and Table~\ref{results_switch_table} indicate using RA-VDN with network in Fig.~\ref{fig:relations}(b), the blue agent accumulates higher individual rewards ($4.40\pm0.00$) as it is the first to use the bridge and than the red agent ($3.80\pm0.00$). Conversely, when the network in Fig.~\ref{fig:relations}(c) was applied, the situation is reversed with the red agent having a higher total reward ($4.39\pm0.02$), as depicted in Fig.~\ref{fig:results_switch}(c). Comparing results from these three experiments reveals that the collective rewards are the same while the relational network influences which \textit{optimal} solution the algorithm converges to.

\vspace{0.3em}
\noindent\textbf{Three-Agent Scenario:}
This setup involves three agents positioned as illustrated in Fig.~\ref{fig:environments}(e). The optimal policy, which minimizes waiting time, is for the pair of agents on the left side to cross the bridge initially, as they are heading in the same direction, either in the order of the red--green--blue or green--red--blue. As illustrated in Fig.~\ref{fig:results_switch}(d) and Table~\ref{results_switch_table}, VDN experiments show that red and green agents have similar individual rewards ($4.38\pm0.04$ and $4.35\pm0.02$) since they randomly alternate taking the lead in passing the bridge over $10$ runs, while the blue agent has lower individual reward ($3.66\pm0.04$). Yet, RA-VDN has the ability to influence the order through the relational network  shown in Fig.~\ref{fig:relations}(e), where the red agent holds the highest significance. Fig.~\ref{fig:results_switch}(e) and Table~\ref{results_switch_table} indicate the bridge usage order becomes consistent as red--green--blue, with individual rewards of $4.40\pm0.00$, $4.30\pm0.00$, and $3.70\pm0.00$, respectively.

Utilizing relational networks can also facilitate the discovery of new behaviors for agents to converge upon, while guiding the algorithm to converge towards an optimal solution. For instance, Fig.~\ref{fig:relations}(f) shows that both the red and green agents prioritize the blue agent, granting it the first turn. Subsequently, the second turn is randomly assigned to either the red or green agent, as their importance weights are equal and they share the same side. The findings of RA-VDN with this network are indicated in Fig.~\ref{fig:results_switch}(b) and Table~\ref{results_switch_table}, which reveal that the blue agent has a higher individual reward ($4.40\pm0.00$) than the others while the red ($3.75\pm0.03$) and green ($3.76\pm0.07$) agents have almost the same rewards due to alternating turns. Furthermore, to assess the agents' adherence to relational networks and cooperative strategies over individual and collective rewards, we introduce a novel relational network illustrated in Fig.~\ref{fig:relations}(g). The network elicits the sequential crossing of the bridge as follows: the green agent crosses from the left side, followed by the blue agent from the right, and the red agent crosses from the left, resulting in the red agent waiting for the alternating turn. Fig.~\ref{fig:results_switch}(g) and Table~\ref{results_switch_table} confirm that the agents prioritize relationships over collective reward, even at a cost. However, the collective reward is lower here as the red agent sacrifices more by taking the last turn, due to the environment's optimal policy being influenced by the relational network, leading to the convergence of the agents towards the \textit{desired} behavior.

\vspace{0.3em}
\noindent\textbf{Four-Agent Scenario:}
In this experimental, four agents are used, with an equal number of agents on each side (Fig.~\ref{fig:environments}(e)). VDN yields a solution that involves stochastic selection of the side that crosses the bridge initially, as well as which agent from that side takes the first turn. The results, depicted in Fig.~\ref{fig:results_switch}(h) and Table~\ref{results_switch_table}, demonstrate that the individual rewards derived from VDN are very similar, owing to the stochasticity induced by the existence of multiple optimal policies that equally weight all the agents. Yet, RA-VDN with the relation network illustrated in Fig.~\ref{fig:relations}(i) has a significant effect on the sequence of bridge usage. The yellow agent is given importance by all agents, leading to the right side taking the initial turn and allowing the yellow agent to be the first to cross the bridge, as shown in Fig.~\ref{fig:results_switch}(i) and Table~\ref{results_switch_table}. We anticipate that the blue agent passes the bridge second to mitigate the undesirable impact of remaining in the environment, as in three-agent scenario. However, due to the increased complexity of the team network and small amount of negative reward, in 80\% of the runs, the blue agent passes the bridge subsequent to the yellow agent, while in the remaining 20\%, the bridge is crossed by either of the three agents randomly. The observed behavior leads to the blue, red, and green agents receiving individual rewards that are highly comparable, while the yellow agent attains the highest reward ($4.34\pm0.09$). 

To address the challenge of the team failing to fully learn the expected team behavior, we propose an extension to our algorithm, which includes freezing the neural network. Initially, the neural network of agents is trained for $1000$ episodes, where the primary agent learns to reach its goal, and the others learn to prioritize it. After the initial phase, we freeze the prioritized agent's network (i.e., its parameters no longer become updated) and it becomes a dynamic element of the environment for subsequent $9000$ episodes of training, with other agents learning their policies accordingly. Through this approach, the complexity of the problem is reduced, and the learning process is accelerated. The results, presented in Fig.~\ref{fig:results_switch}(j) and Table~\ref{results_switch_table}, demonstrate that the yellow agent achieves the highest reward ($4.40\pm0.00$) among the four agents because it is the initial agent to utilize the bridge. The blue agent's reward ($4.20\pm0.00$) surpasses that of the remaining agents because it can pass the bridge right after the yellow agent, while the individual rewards of the green ($3.54\pm0.14$) and red ($3.47\pm0.15$) agents are close due to the alternate turns they take.

\section{Conclusion and Future Work}
We propose a novel framework that incorporates relationship awareness into agents' learning algorithms. Our framework enables the discovery of new cooperative behaviors and the ability to guide agents' behavior based on inter-agent relationships. Our experiments, conducted in a multi-agent grid-world environment, validate the effectiveness of our approach in solving environments with agents having different abilities. Moreover, in the \textit{Switch} environment, we demonstrated the algorithm's power in influencing agents' behavior within a team. However, we acknowledge that adjusting relational weights can become challenging as the relational network becomes denser with an increasing number of agents. To overcome this issue, we introduce the idea of freezing the agents neural networks, which we show to be effective in reducing problem complexity and expediting the learning process. As a next step, we aim to conduct additional experiments in more complex environments that involve an increased number of agents and compare the performance of our algorithm with other state-of-the-art methods.

\section{Acknowledgments}

This work is supported in part by NSF (IIS-2112633) and
the Army Research Lab (W911NF20-2-0089).

\bibliographystyle{IEEEtran}
\bibliography{IEEEabrv, refs}

\begin{thebibliography}{10}
\providecommand{\url}[1]{#1}
\csname url@rmstyle\endcsname
\providecommand{\newblock}{\relax}
\providecommand{\bibinfo}[2]{#2}
\providecommand\BIBentrySTDinterwordspacing{\spaceskip=0pt\relax}
\providecommand\BIBentryALTinterwordstretchfactor{4}
\providecommand\BIBentryALTinterwordspacing{\spaceskip=\fontdimen2\font plus
\BIBentryALTinterwordstretchfactor\fontdimen3\font minus
  \fontdimen4\font\relax}
\providecommand\BIBforeignlanguage[2]{{%
\expandafter\ifx\csname l@#1\endcsname\relax
\typeout{** WARNING: IEEEtran.bst: No hyphenation pattern has been}%
\typeout{** loaded for the language `#1'. Using the pattern for}%
\typeout{** the default language instead.}%
\else
\language=\csname l@#1\endcsname
\fi
#2}}

\bibitem{kleiner2006rfid}
A.~Kleiner, J.~Prediger, and B.~Nebel, ``Rfid technology-based exploration and
  slam for search and rescue,'' in \emph{2006 IEEE/RSJ International Conference
  on Intelligent Robots and Systems}.\hskip 1em plus 0.5em minus 0.4em\relax
  IEEE, 2006, p. 4054.

\bibitem{pendleton2017perception}
S.~D. Pendleton, H.~Andersen, X.~Du, X.~Shen, M.~Meghjani, Y.~H. Eng, D.~Rus,
  and M.~H. Ang~Jr, ``Perception, planning, control, and coordination for
  autonomous vehicles,'' \emph{Machines}, vol.~5, no.~1, 2017.

\bibitem{Kim2022CACC}
T.~Kim and K.~Jerath, ``Congestion-aware cooperative adaptive cruise control
  for mitigation of self-organized traffic jams,'' \emph{IEEE Transactions on
  Intelligent Transportation Systems}, vol.~23, no.~7, pp. 6621--6632, 2022.

\bibitem{peng2017multiagent}
P.~Peng, Y.~Wen, Y.~Yang, Q.~Yuan, Z.~Tang, H.~Long, and J.~Wang, ``Multiagent
  bidirectionally-coordinated nets: Emergence of human-level coordination in
  learning to play starcraft combat games,'' \emph{arXiv preprint
  arXiv:1703.10069}, 2017.

\bibitem{busoniu2008comprehensive}
L.~Busoniu, R.~Babuska, and B.~De~Schutter, ``A comprehensive survey of
  multiagent reinforcement learning,'' \emph{IEEE Transactions on Systems, Man,
  and Cybernetics, Part C (Applications and Reviews)}, vol.~38, no.~2, pp.
  156--172, 2008.

\bibitem{shoham2007if}
Y.~Shoham, R.~Powers, and T.~Grenager, ``If multi-agent learning is the answer,
  what is the question?'' \emph{Artificial intelligence}, vol. 171, no.~7, pp.
  365--377, 2007.

\bibitem{Haeri2020Swarm}
\BIBentryALTinterwordspacing
H.~Haeri, K.~Jerath, and J.~Leachman, ``{Thermodynamics-Inspired Macroscopic
  States of Bounded Swarms},'' \emph{ASME Letters in Dynamic Systems and
  Control}, vol.~1, no.~1, 03 2020, 011015. [Online]. Available:
  \url{https://doi.org/10.1115/1.4046580}
\BIBentrySTDinterwordspacing

\bibitem{matignon2012independent}
L.~Matignon, G.~J. Laurent, and N.~Le~Fort-Piat, ``Independent reinforcement
  learners in cooperative markov games: a survey regarding coordination
  problems,'' \emph{The Knowledge Engineering Review}, vol.~27, no.~1, pp.
  1--31, 2012.

\bibitem{oliehoek2008optimal}
F.~A. Oliehoek, M.~T. Spaan, and N.~Vlassis, ``Optimal and approximate q-value
  functions for decentralized pomdps,'' \emph{Journal of Artificial
  Intelligence Research}, vol.~32, pp. 289--353, 2008.

\bibitem{gronauer2022multi}
S.~Gronauer and K.~Diepold, ``Multi-agent deep reinforcement learning: a
  survey,'' \emph{Artificial Intelligence Review}, pp. 1--49, 2022.

\bibitem{guestrin2001multiagent}
C.~Guestrin, D.~Koller, and R.~Parr, ``Multiagent planning with factored
  mdps,'' \emph{Advances in neural information processing systems}, vol.~14,
  2001.

\bibitem{bohmer2020deep}
W.~B{\"o}hmer, V.~Kurin, and S.~Whiteson, ``Deep coordination graphs,'' in
  \emph{International Conference on Machine Learning}.\hskip 1em plus 0.5em
  minus 0.4em\relax PMLR, 2020, pp. 980--991.

\bibitem{sunehag2017value}
P.~Sunehag, G.~Lever, A.~Gruslys, W.~M. Czarnecki, V.~Zambaldi, M.~Jaderberg,
  M.~Lanctot, N.~Sonnerat, J.~Z. Leibo, K.~Tuyls, \emph{et~al.},
  ``Value-decomposition networks for cooperative multi-agent learning,''
  \emph{arXiv preprint arXiv:1706.05296}, 2017.

\bibitem{claus1998dynamics}
C.~Claus and C.~Boutilier, ``The dynamics of reinforcement learning in
  cooperative multiagent systems,'' \emph{AAAI/IAAI}, vol. 1998, no. 746-752,
  p.~2, 1998.

\bibitem{tan1993multi}
M.~Tan, ``Multi-agent reinforcement learning: Independent vs. cooperative
  agents,'' in \emph{Proceedings of the tenth international conference on
  machine learning}, 1993, pp. 330--337.

\bibitem{watkins1992q}
C.~J. Watkins and P.~Dayan, ``Q-learning,'' \emph{Machine learning}, vol.~8,
  pp. 279--292, 1992.

\bibitem{tampuu2017multiagent}
A.~Tampuu, T.~Matiisen, D.~Kodelja, I.~Kuzovkin, K.~Korjus, J.~Aru, J.~Aru, and
  R.~Vicente, ``Multiagent cooperation and competition with deep reinforcement
  learning,'' \emph{PloS one}, vol.~12, no.~4, 2017.

\bibitem{hernandez2017survey}
P.~Hernandez-Leal, M.~Kaisers, T.~Baarslag, and E.~M. De~Cote, ``A survey of
  learning in multiagent environments: Dealing with non-stationarity,''
  \emph{arXiv preprint arXiv:1707.09183}, 2017.

\bibitem{zhang2018fully}
K.~Zhang, Z.~Yang, H.~Liu, T.~Zhang, and T.~Basar, ``Fully decentralized
  multi-agent reinforcement learning with networked agents,'' in
  \emph{International Conference on Machine Learning}.\hskip 1em plus 0.5em
  minus 0.4em\relax PMLR, 2018, pp. 5872--5881.

\bibitem{lowe2017multi}
R.~Lowe, Y.~I. Wu, A.~Tamar, J.~Harb, O.~Pieter~Abbeel, and I.~Mordatch,
  ``Multi-agent actor-critic for mixed cooperative-competitive environments,''
  \emph{Advances in neural information processing systems}, vol.~30, 2017.

\bibitem{yu2021surprising}
C.~Yu, A.~Velu, E.~Vinitsky, Y.~Wang, A.~Bayen, and Y.~Wu, ``The surprising
  effectiveness of ppo in cooperative, multi-agent games,'' \emph{arXiv
  preprint arXiv:2103.01955}, 2021.

\bibitem{rashid2020monotonic}
T.~Rashid, M.~Samvelyan, C.~S. De~Witt, G.~Farquhar, J.~Foerster, and
  S.~Whiteson, ``Monotonic value function factorisation for deep multi-agent
  reinforcement learning,'' \emph{The Journal of Machine Learning Research},
  vol.~21, no.~1, pp. 7234--7284, 2020.

\bibitem{son2019qtran}
K.~Son, D.~Kim, W.~J. Kang, D.~E. Hostallero, and Y.~Yi, ``Qtran: Learning to
  factorize with transformation for cooperative multi-agent reinforcement
  learning,'' in \emph{International conference on machine learning}.\hskip 1em
  plus 0.5em minus 0.4em\relax PMLR, 2019, pp. 5887--5896.

\bibitem{baker2020emergent}
B.~Baker, ``Emergent reciprocity and team formation from randomized uncertain
  social preferences,'' \emph{Advances in neural information processing
  systems}, vol.~33, pp. 15\,786--15\,799, 2020.

\bibitem{haeri2022reward}
H.~Haeri, R.~Ahmadzadeh, and K.~Jerath, ``Reward-sharing relational networks in
  multi-agent reinforcement learning as a framework for emergent behavior,''
  \emph{arXiv preprint arXiv:2207.05886}, 2022.

\bibitem{mnih2015human}
V.~Mnih, K.~Kavukcuoglu, D.~Silver, A.~A. Rusu, J.~Veness, M.~G. Bellemare,
  A.~Graves, M.~Riedmiller, A.~K. Fidjeland, G.~Ostrovski, \emph{et~al.},
  ``Human-level control through deep reinforcement learning,'' \emph{nature},
  vol. 518, no. 7540, pp. 529--533, 2015.

\bibitem{magym}
A.~Koul, ``ma-gym: Collection of multi-agent environments based on openai
  gym.'' \url{https://github.com/koulanurag/ma-gym}, 2019.

\end{thebibliography}

\end{document}